# Intrinsic water transport in moisture-capturing hydrogels

*Gustav Graeber[1,2,*,#], Carlos D. Díaz-Marín[1,#] and Leon C. Gaugler[1]*

[1]Device Research Laboratory, Department of Mechanical Engineering, MIT

[2]Graeber Lab for Energy Research, Chemistry Department, Humboldt Universität zu Berlin

[#]Equal contribution

*gustav.graeber@hu-berlin.de

**Abstract**

Moisture-capturing hydrogels have emerged as attractive sorbent materials capable of converting ambient humidity into liquid water. Recent works have demonstrated exceptional water capture capabilities of hydrogels, while simultaneously, exploring different strategies to accelerate water capture and release. However, on the material level, an understanding of the intrinsic transport properties of moisture-capturing hydrogels is currently missing, which hinders their rational design. In this work, we combine absorption and desorption experiments of macroscopic hydrogel samples in pure-vapor with models of water diffusion in the hydrogels to demonstrate the first measurements of the intrinsic water diffusion coefficient in hydrogel-salt composites. Based on these insights, we pattern hydrogels with micropores to significantly decrease the required absorption and desorption time by 19% and 72%, respectively, while reducing the total water capacity of the hydrogel by only 4%. Thereby, we provide an effective strategy towards hydrogel material optimization, with a particular significance in pure-vapor environments.

Water sorption from the air is ubiquitous in nature and relevant for applications such as atmospheric freshwater harvesting,[1–5] passive cooling,[6–9] and thermal energy storage.[5,10–12] Composites from moisture-capturing salts embedded into a hydrogel matrix have recently emerged as exceptional materials for sorption, exhibiting low cost, high scalability, and large ability to capture and retain water from the air, i.e., high water uptake.[13,14] Recent efforts have attempted to increase the absorption and desorption speed of hydrogels, i.e., kinetics, which is critical towards high-performance sorption. These works have relied on strategies such as reduction of material thickness,[14,15] fabrication of low tortuosity porous hydrogels,[16] and utilization of polyelectrolyte gels,[17] thereby achieving fast hydrogel absorption and desorption speeds. However, the mechanisms behind this enhancement can limit the overall application performance or have not been rigorously demonstrated. For instance, strategies that reduce the characteristic material length or introduce excessive porosity will reduce the overall water that can be absorbed into the hydrogel. On the other hand, mechanisms behind intrinsic material-level enhancements[17] have not been clarified and connected with the mass transport processes occurring inside the hydrogel.[18,19] Mechanistic insights are hindered by the fact that current measurements of hydrogel kinetics rely on macroscopic samples. Therefore, these investigations are uncapable of decoupling system-level properties (such as convection), geometric properties (such as thickness and morphology), and intrinsic hydrogel transport properties (such as diffusion inside the hydrogel).[14,17] On the other hand, dynamic vapor sorption measurements, which are conducted in a pure-vapor environment, can isolate intrinsic material properties. However, these measurements require milligram-scale samples, which hinder the control of macroscopic geometrical features and the interpretation of the results in terms of transport properties.

In this work, for the first time, we probe the intrinsic transport processes, such as diffusion and crystallization, occurring in a hydrogel during water absorption and desorption. We perform sorption experiments in a custom-built pure-vapor chamber and mass transfer modeling to isolate the transport properties within the hydrogel and measure an effective diffusion coefficient of ≈ $1.8 \cdot 10^{-10}$ m$^2$ s$^{-1}$ for both absorption and desorption, consistent with previous models.[18,19] Furthermore, we combine a simple one-pot synthesis with micropatterning of hydrogels towards the improvement of kinetics by reducing the effective diffusion resistance. In particular, by designing the porosity and the pore radius of the hydrogels, we reduce the desorption and absorption timescales by 19% and 72%, respectively, while still retaining 96% of the moisture capture capability of a nonporous sample. Our results and modeling further show key features of absorption and desorption in pure-vapor environments, such as a pronounced effect of non-condensable gases, which reduce the kinetics especially during absorption. These results, represent a significant step towards the understanding of the processes governing water absorption and desorption by hydrogels, which is critical towards application-specific optimization. Furthermore, the strategy demonstrated here represents a key enhancement towards moisture-capturing hydrogels with high water uptake and fast kinetics in pure-vapor environments, which can lead to optimized sorption-based thermal energy storage.

In **Figure 1**, we summarize the synthesis, characterization and water uptake properties of the lithium chloride-loaded polyacrylamide (PAM) hydrogel composites that we prepared in this work, see S1 for details. Notably, we used a one-pot synthesis where polymerization occurs in a salt solution, enabling a scalable synthesis with a controlled salt content. **Figure 1a** shows one sample in its dry state. As water evaporated, the salt crystalized, leading to a solid, white sample. As the samples were exposed to ambient humidity, they captured water from the air until the salt deliquesces. The samples then recovered their gel-like properties (**Figure 1b**). With our one-pot synthesis, we could easily control the salt loading, and, therefore, the hydrogel moisture-capturing capabilities. For instance, we synthesized hydrogel samples with 0, 1, and 5 gram of LiCl per gram of acrylamide (AM), which we denote as $g_{LiCl}\ g_{AM}^{-1}$.

After synthesis, we characterized the water uptake of the hydrogel-salt composites using a dynamic vapor sorption (DVS) apparatus. In **Figure 1c**, we show the equilibrium uptake of PAM hydrogels as a function of relative vapor pressure (RP) and LiCl content. Since the sample and environment are isothermal during our measurements, RP is equivalent to the relative humidity. The water uptake is defined as the mass of captured water vapor divided by the mass of the dry, water-free sample, which we denote as $g_{H2O}\ g_{dry}^{-1}$. **Figure 1c** shows the resulting isotherms for the three different compositions, namely 0 $g_{LiCl}\ g_{AM}^{-1}$, 1 $g_{LiCl}\ g_{AM}^{-1}$, and 5 $g_{LiCl}\ g_{AM}^{-1}$. The water uptake of the salt-free sample (0 $g_{LiCl}\ g_{AM}^{-1}$) is practically 0 up to 60% RP and reaches very low values of 0.31 $g_{H2O}\ g_{dry}^{-1}$ at 90% RP. In contrast, the 1 $g_{LiCl}\ g_{AM}^{-1}$ and the 5 $g_{LiCl}\ g_{AM}^{-1}$ sample show a substantially higher ability to capture vapor. E.g., at 30%, 60%, and 90% RP, the 5 $g_{LiCl}\ g_{AM}^{-1}$ sample captures 1.41, 2.46, and 6.67 $g_{H2O}\ g_{dry}^{-1}$, respectively. In this work, we focused on the 5 $g_{LiCl}\ g_{AM}^{-1}$ samples, as their water uptakes are comparable to the highest reported values for hygroscopic hydrogels.[13,16,20,21]

The previous measurements are limited to milligram-scale samples due to typical constraints with DVS machines. These constraints hinder the use of macroscopic and well-defined samples, which are necessary for measurements of material-level kinetics. Therefore, previous large-scale sample measurements were performed outside of DVS machines in air-vapor environments. However, in air-vapor environments, convection of air from the ambient to the hydrogel significantly slows down the hydrogel kinetics, thereby not allowing to isolate the intrinsic material properties.

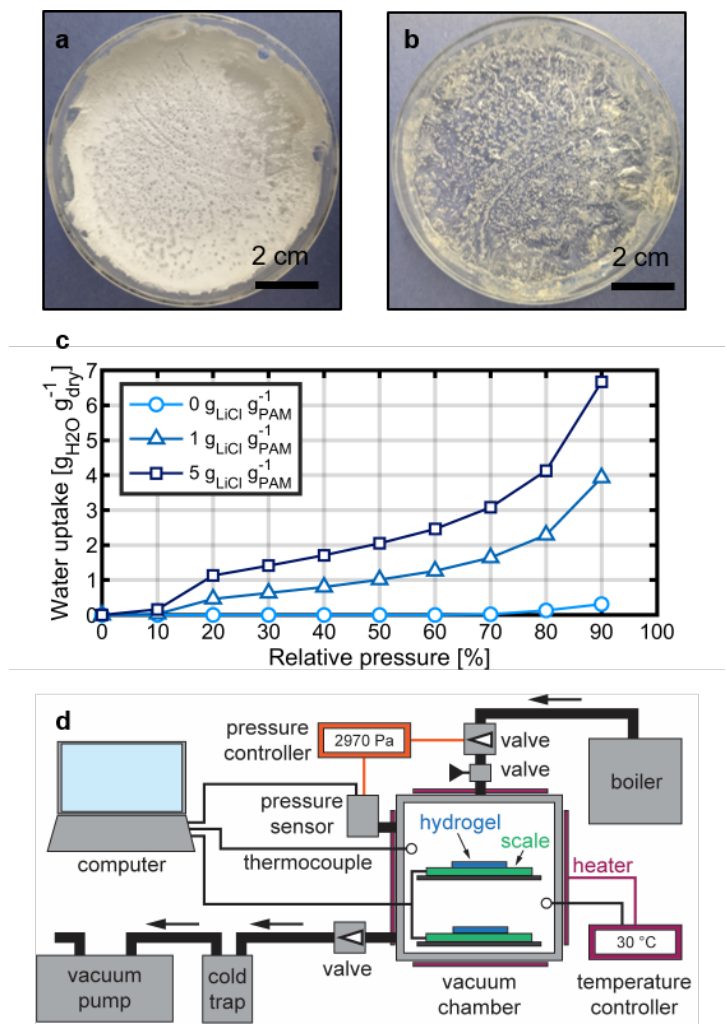

**Figure 1. Synthesis and characterization of hydrogel-salt composites. a**, Dried PAM hydrogel containing 5 g of LiCl per g of monomer. The dried sample was solid and white due to the large salt content. **b**, PAM hydrogel with 5 g of LiCl per g of monomer after being exposed to ambient moisture and recovering its gel-like properties. **c**, Dynamic vapor sorption uptake measurements of milligram-scale hydrogel samples with varying salt content. PAM with no salt exhibits almost no uptake. The uptake is significantly increased when 5 g of LiCl are added per g of AM. **d**, Schematic of the experimental setup to study hydrogel sorption and desorption kinetics in a pure-vapor environment.

It is therefore desirable to study macroscopic samples in pure-vapor environments, where the dominant transport resistance originates from within the materials. To achieve this, we built an experimental setup to study hydrogel sorption and desorption kinetics in a pure-vapor environment, **Figure 1d**. The setup consists of a highly leak-tight vacuum chamber with two data-logging scales to simultaneously monitor the weight of two hydrogel samples as a function of the environmental conditions. The chamber is connected to a vacuum pump to reduce RP and to a boiler to add more vapor to the system. The chamber is equipped with heaters to generate an environment with a steady temperature of 30 °C. Pressure and temperature sensors are used to control the conditions inside the chamber. A photograph and further details about the experimental procedures are provided in S2. The system that we developed combines the pure-vapor advantages of a DVS system, while also enabling the study of samples of up to 600 g in mass, exceeding the capabilities of DVS systems by several orders of magnitude. This unique setup allowed us to probe the intrinsic transport properties of hydrogel-salt composites.

**Figure 2a and b** show the uptake of a 5 $g_{LiCl}\ g_{AM}^{-1}$ hydrogel with a thickness of 9 mm as a function of time for desorption and absorption, respectively. Initially, the samples were placed in the as-fabricated condition in the environmental chamber. Then, the pressure was reduced to near vacuum conditions (~0% RP). Due to the low RP, water desorbed from the sample until after 30 hours the water uptake was reduced to ~1 $g_{H2O}\ g_{dry}^{-1}$, which is approximately half of the initial value (**Figure 2a**). Before the start of the absorption step shown in **Figure 2b**, the sample was exposed for over two weeks to a vacuum on the order of 10 Pa at a temperature of 30 °C until it reached an uptake 0.25 $g_{H2O}\ g_{dry}^{-1}$. Subsequently, at time zero in **Figure 2b**, the sample was exposed to a vapor pressure of around 2970 Pa, i.e., 70% RP. Under these high RP conditions, the dry hydrogel started to absorb water from the pure-vapor environment and its weight increased. After less than 30 hours, the water uptake had increased in over 8 times relative to its initial value to exceed 2 $g_{H2O}\ g_{dry}^{-1}$.

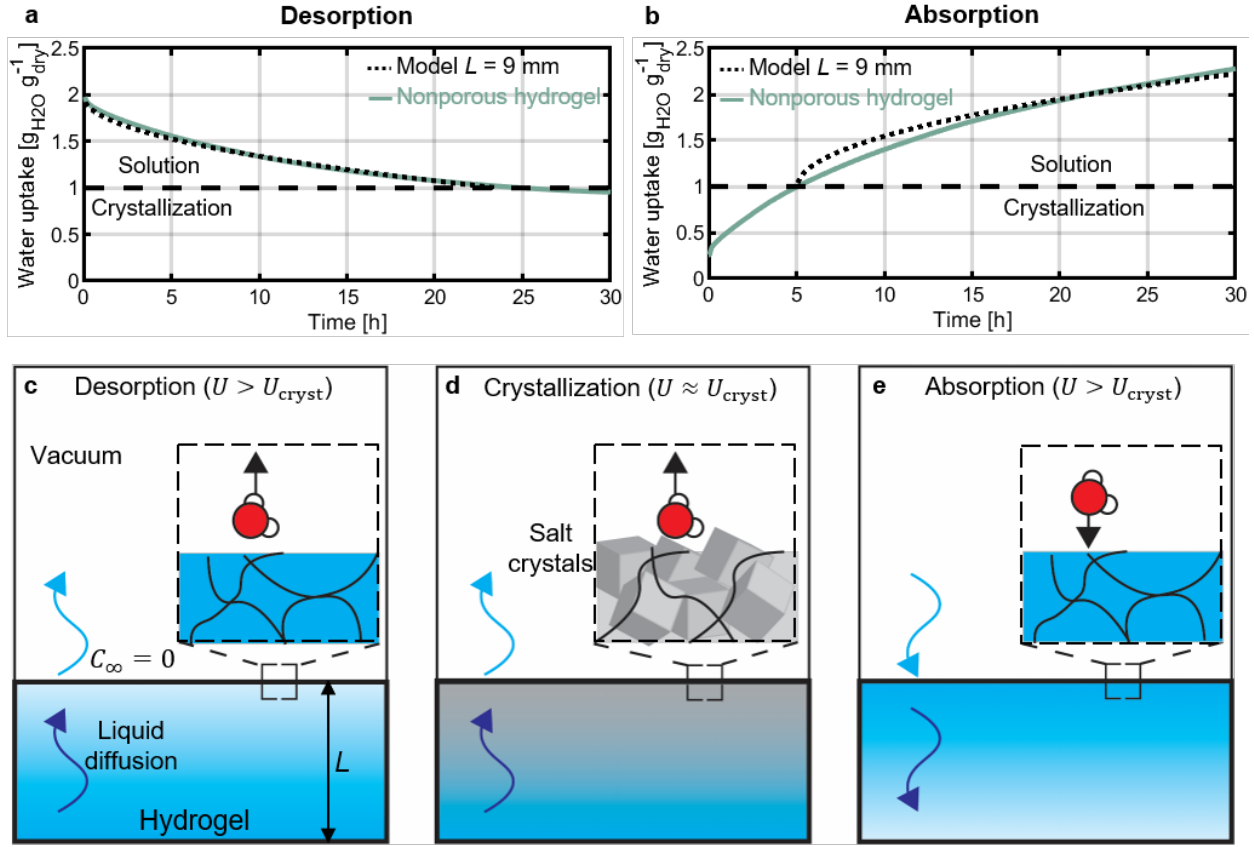

**Figure 2. Desorption and absorption of hydrogel-salt composites. a**, Desorption of nonporous LiCl-loaded PAM hydrogel. The salt is initially dissolved in the water in the hydrogel. As desorption progresses beyond the crystallization point with an uptake of less than 1 $g_{H2O}\ g_{dry}^{-1}$, the salt crystallizes. **b**, Absorption of nonporous LiCl-loaded PAM hydrogel. The salts start from a crystallized state and capture water vapor until they deliquesce. For both desorption and absorption, the model capturing the diffusion of water inside the hydrogels agrees well with experiments. This agreement enables us to measure the effective diffusion coefficient. **c**, Schematic illustration of desorption during the solution-regime, where the water uptake of the hydrogel, $U$, is higher than that at crystallization. At the free interface of the hydrogel, the water concentration is fixed at $c_\infty = 0$ and water molecules escape from the hydrogel to the vacuum. Inside the hydrogel, water diffuses to the interface across a diffusion length equal to the thickness of the hydrogel. **d**, Once the salt crystallizes, desorption proceeds through a different mechanism. Parts of the hydrogel contain crystallized salt hydrates and further desorption proceeds by either growth of the crystallized region or desorption of water molecules from the hydrates. **e**, During absorption, mass transfer is reversed relative to desorption. Specifically, once the salts have deliquesced, water molecules condense from the ambient at the interface. Water is then diffused inside the hydrogel.

During this desorption and absorption cycle, different physical phenomena occur. Initially, during desorption, water leaves the surface of the hydrogel driven by the vapor pressure difference between the surface and the environment. The water concentration at the top of the hydrogel decreases, leading to a diffusive flux of water towards the hydrogel surface (**Figure 2c**). The water concentration at the free surface of the hydrogel continues to decrease, until it eventually reaches the solubility limit of lithium chloride in water. At this point, the salt at the surface crystallizes into hydrates (**Figure 2d**). Further desorption leads to a larger volume of the hydrogel having salt hydrates. We estimate the water uptake at which the sample crystallizes as $U_{cryst} = 1$ $g_{H2O}$ $g_{dry}^{-1}$ (see S3 for details). To measure the intrinsic diffusion coefficient of water in the hydrogel, we focus on desorption before crystallization. For this regime, we calculate the desorption uptake, $U_{des}$, as a function of time, $t$, as (see S4 for derivation)

$$U_{des}(t) = \frac{2U_0}{\pi^2} \sum_{n=0}^{\infty} \frac{\exp\left(-\left(n+\frac{1}{2}\right)^2 \pi^2 D_{des} t/L^2\right)}{\left(n+\frac{1}{2}\right)^2}, \tag{1}$$

where $U_0$ is the initial uptake, given by the as-synthesized condition in our experiments, $D_{des}$ is the effective diffusion coefficient of water in the hydrogel, and $L$ is the characteristic diffusion length, corresponding here to the thickness of the hydrogel sample. **Figure 2a** compares the results from Equation (1) with our experiments, showing good agreement between both. Specifically, we fitted Equation (1) to our experimental results using $D_{des}$ as a fitting parameter, yielding a measured value $D_{des} = 1.84 \times 10^{-10}$ m² s⁻¹. This approach to measure the diffusion coefficient mirrors commonly used measurement techniques of diffusivity in nano-porous solids.[22]

During absorption, the physical processes are reversed relative to desorption. First, for a crystallized sample, vapor absorption increases the water concentration at the top of the sample, until it reaches the solubility limit of lithium chloride in water. At this point, the salt locally deliquesces. The volume of the hydrogel with salt solution continues to increase as absorption continues, until all the salt has deliquesced. At this point, absorption consists of simultaneous condensation of water molecules at the hydrogel surface and downwards diffusion of water. The absorption uptake during this process, $U_{\text{abs}}$, is given by

$$U_{\text{abs}}(t) = U_\infty + \frac{2(U_0 - U_\infty)}{\pi^2} \left( \sum_{n=0}^{\infty} \frac{\exp\left(-\left(n + \frac{1}{2}\right)^2 \pi^2 D_{\text{abs}} t / L^2\right)}{\left(n + \frac{1}{2}\right)^2} \right), \quad (2)$$

where $U_\infty$ is the equilibrium uptake, corresponding to 3.43 $g_{H2O}$ $g_{dry}^{-1}$ at a RP of 70% (see S4 for details), $U_0$ is the initial uptake, and $D_{\text{abs}}$ is the effective diffusion coefficient of water during absorption. We considered $U_0 = U_{\text{cryst}}$, since we aim to model absorption after deliquescence. **Figure 1b** compares the results from Equation (2) with our experiments, showing good agreement. From fitting our model to the experiments, we calculate $D_{\text{abs}} = 1.81 \times 10^{-10}$ m² s⁻¹. Based on our absorption and desorption experiments, our measured effective diffusion coefficient, $D$, is $D_{\text{abs}} \approx D_{\text{des}} \approx 1.8 \times 10^{-10}$ m² s⁻¹, where the similar values between absorption and desorption validate our measurements and model. The similar diffusion coefficients reflect the ability of our approach to actually isolate the intrinsic transport properties of the material. This is in contrast with previous measurements yielding higher desorption diffusion coefficients,[23] resulting from system-level and convective effects. Furthermore, this measured diffusion coefficient is in agreement with previous model estimates based on poroelastic transport in hydrogels.[18,19] We also note that previous works have reported sorption timescales for their materials.[14,15] These, however, are dependent on the sample geometry which prevents from comparing the material properties between different works.

Our work overcomes this limitation, by measuring diffusion coefficients which are independent of the sample size and morphology.

Our pure-vapor setup can also be used to study the effect of hydrogel porosity and pore radii on the absorption and desorption kinetics. By introducing pores, the water diffusion distance in the hydrogel is reduced to less than the thickness of the hydrogel. For a porous sample with a regular pattern as studied here, this effective diffusion distance, $L_{\text{eff}}$, can be considered as half the closest distance between the pores exposed to the pure vapor environment (**Figure 3a**). This is unique to our pure vapor measurements, as conventional measurements in air-vapor environments do not have a constant vapor concentration along the entire hydrogel surface. In contrast, in an air-vapor environment, vapor has to diffuse into or out of the pores and a concentration gradient will be present.[18,19]

The total amount of water that can be absorbed per hydrogel footprint area, $M$, and the corresponding water absorption rate, $\dot{M}$, will be different for a porous hydrogel compared to a nonporous one (**Figure 3**). These two variables are given by

$$M = U(1-\phi)H_0\rho_{\text{hyd,dry}}, \tag{3}$$

$$\dot{M} = \frac{U(1-\phi)H_0\rho_{\text{hyd,dry}}}{\frac{L_{\text{eff}}^2}{D}}, \tag{4}$$

where $H_0$ and $\rho_{\text{hyd,dry}}$ are the hydrogel thickness and density in the completely dry state, $L_{\text{eff}}$ is the effective diffusion length, and $D$ is the diffusion coefficient of water in the hydrogel. We note that in defining Equation (4), we have considered that $\dot{M}$ is the average absorption rate and that over a timescale of ($L_{\text{eff}}^2/D$) the system reaches equilibrium (see S5 for justification).

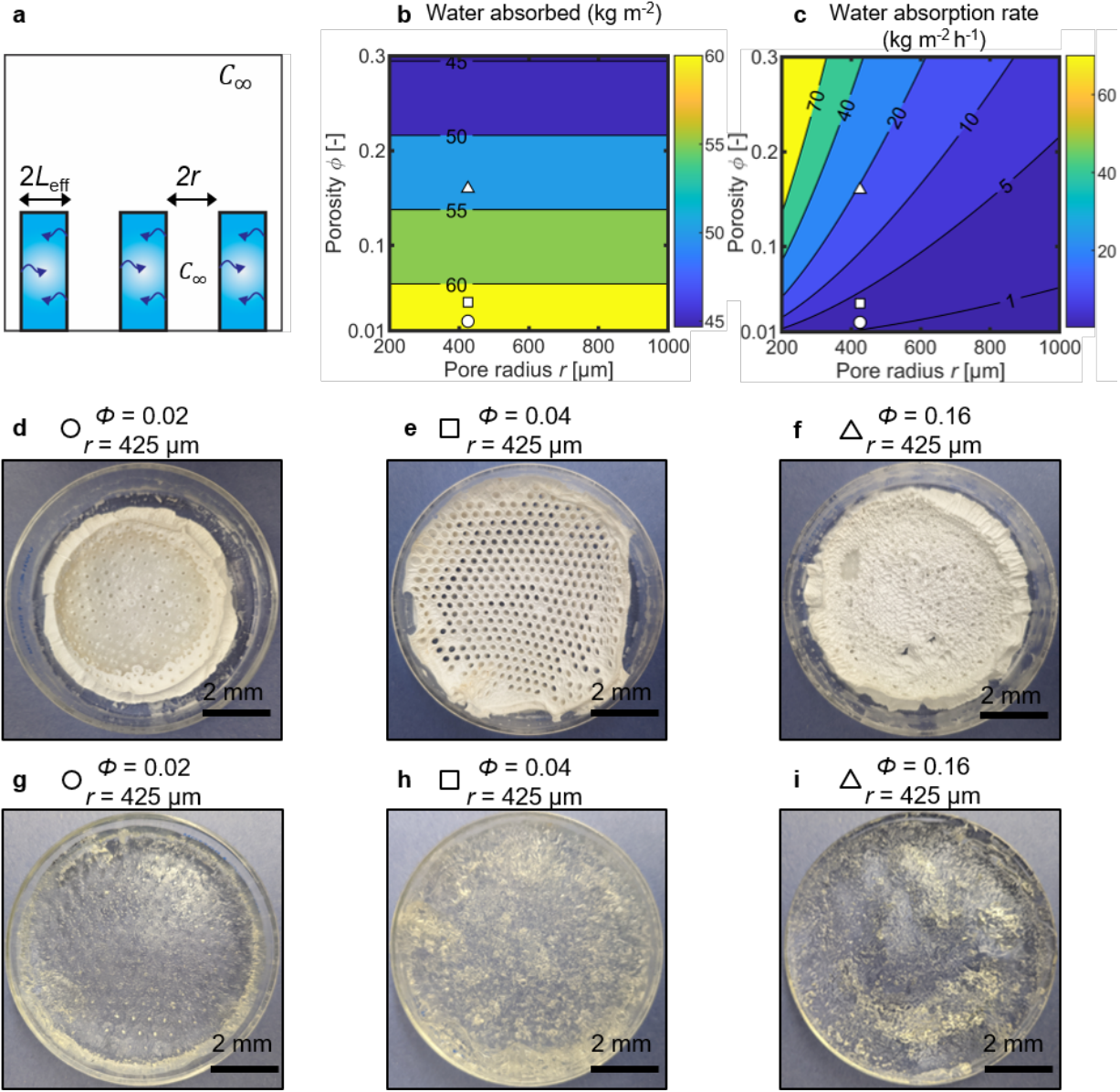

**Figure 3. Design and synthesis of porous hydrogels. a**, In a pure-vapor environment, patterning pores with a radius, $r$, can significantly reduce the effective diffusion length, $L_{\text{eff}}$, and thereby speed up absorption and desorption. Transport resistances in the vapor are negligible, since the same concentration, $C_\infty$, will be present everywhere and also inside the pores. **b**, Water absorbed per footprint area as a function of pore radius, $r$, and porosity, $\phi$, for a 9 mm thick hydrogel with 5 $g_{\text{LiCl}}$ $g_{\text{AM}}^{-1}$ exposed to 70% RP. Increasing the porosity monotonically decreases the water that can be absorbed due to loss of hydrogel material. **c**, Water absorption rate as a function of $r$ and $\phi$ for a hydrogel with the same parameters as those considered in Figure 3b and with $D = 1 \times 10^{-10}$ m$^2$ s$^{-1}$. Based on the modeling, we synthesized porous hydrogels with controlled porosity and pore size using a molding process. We chose the minimum pore radius compatible with our molding process (425 μm) and fabricated three samples with varying porosity of, **d**, 0.02; **e**, 0.04; and, **f**, 0.16,

shown in their dry state. When exposing the samples to ambient humidity, the hydrogels regained their gel-like properties as shown in **g**, **h**, and **i**. The most porous samples exhibited significant deformation as they captured vapor.

Equations (3) and (4) illustrate a critical point, showing that by introducing porosity, the total water captured is reduced, but the capture rate can be increased due to the smaller diffusion resistances. **Figure 3b** shows the impact of the porosity and the pore radius on $M$ for a representative system with $U$ = 3.43 $g_{H2O}$ $g_{dry}^{-1}$ (corresponding to a hydrogel at 70% RP and with 5 $g_{LiCl}$ $g_{AM}^{-1}$), $H_0 = 9$ mm, and $\rho_{hyd,dry}$ (approximating the density to that of dry lithium chloride[24]). As captured by Equation (3), increasing the porosity monotonically decreases the total water captured. For instance, increasing the porosity from 0 to 0.3 decreases the absorbed water from ≈60 kg m$^{-2}$ to ≈45 kg m$^{-2}$. The pore radius, which was considered independent of porosity, has no effect on the total amount of captured water. **Figure 3c** shows the effect of porosity and pore radius on $\dot{M}$ (see S6 for $L_{eff}$ as a function of $\phi$ and $r$ for the considered geometry). Increasing the porosity from 0 to 0.3 monotonically increases the absorption rate, driven by a sizeable decrease in $L_{eff}$. On the other hand, reducing the pore size while maintaining a constant porosity monotonically increases the absorption rate as a result of smaller diffusion lengths.

    We leveraged the previous modeling insights to design porous hydrogels. Specifically, we fabricated porous hydrogels with well-defined pores and interpore distance using custom-made PDMS molds with micropores at controlled distances (see Figure S3). As a result, we could independently control $r$ and $\phi$. Additionally, combining the molding with the one-pot hydrogel synthesis, our method overcame limits of scalability, equipment requirements, and time-consumption suffered by typical approaches relying on freeze-drying and swelling.[7,13,14,25] Guided by our model, we selected $r$ = 425 μm, which was the smallest radius achievable with our

fabrication. We fabricated samples with varying pitches to explore the limits of porosity achievable with our method. **Figures 3d, e,** and **f** show samples with $\phi$ = 0.02, 0.04, and 0.16, respectively, in their dry state. We note that as the sample with the highest porosity was dried, it significantly lost its structural integrity, also losing its well-defined porosity. In contrast, the samples with the lower porosity retained their pores after drying. **Figures 3 g, h,** and **i** show the same samples in their hydrated state after being exposed to ambient humidity for over three days. We note that especially for the highest porosity sample, the low rigidity led to higher sample deformation in the hydrated state. Based on these findings, we focused on the sample with the intermediate porosity, as it exhibited stability and the potential for an increase of more than eleven-fold in the kinetics (see S6) with only a minor reduction of the water absorbed of ~4%.

With the selected porous sample, we performed desorption and absorption experiments identical to those carried out with the nonporous samples. **Figure 4a** compares the uptake during desorption of the porous sample with the nonporous sample. The porous sample desorbs water significantly faster, reaching the crystallization uptake in ~7 hours, compared to the ~25 hours required by the nonporous sample. Furthermore, we can leverage the previous model of Equation (3) to obtain an experimental effective diffusion length. Specifically, we considered $D_{des} = 1.84 \times 10^{-10}$ m² s⁻¹ (as obtained from the desorption of the nonporous sample) and fitted the experimental results to Equation (3) to obtain $L_{eff}$ = 4.8 mm. We highlight that this value is smaller than the overall sample thickness of 9 mm, reflecting the success of our approach in reducing the diffusion length. It is also comparable to the value expected from our design, i.e., 2.6 mm, where we attribute the higher experimental value to defects present in our molding process and reduced porosity near the sample edges (**Figure 3e**). We can further observe the effect of the pores on the kinetics by comparing the desorption rate (**Figure 4b**). Initially, the porous sample has a desorption rate with a significantly higher magnitude. However, the uptake approaches crystallization, the desorption rates of both samples are similar, reflecting the non-diffusive nature of transport once crystallization occurs.

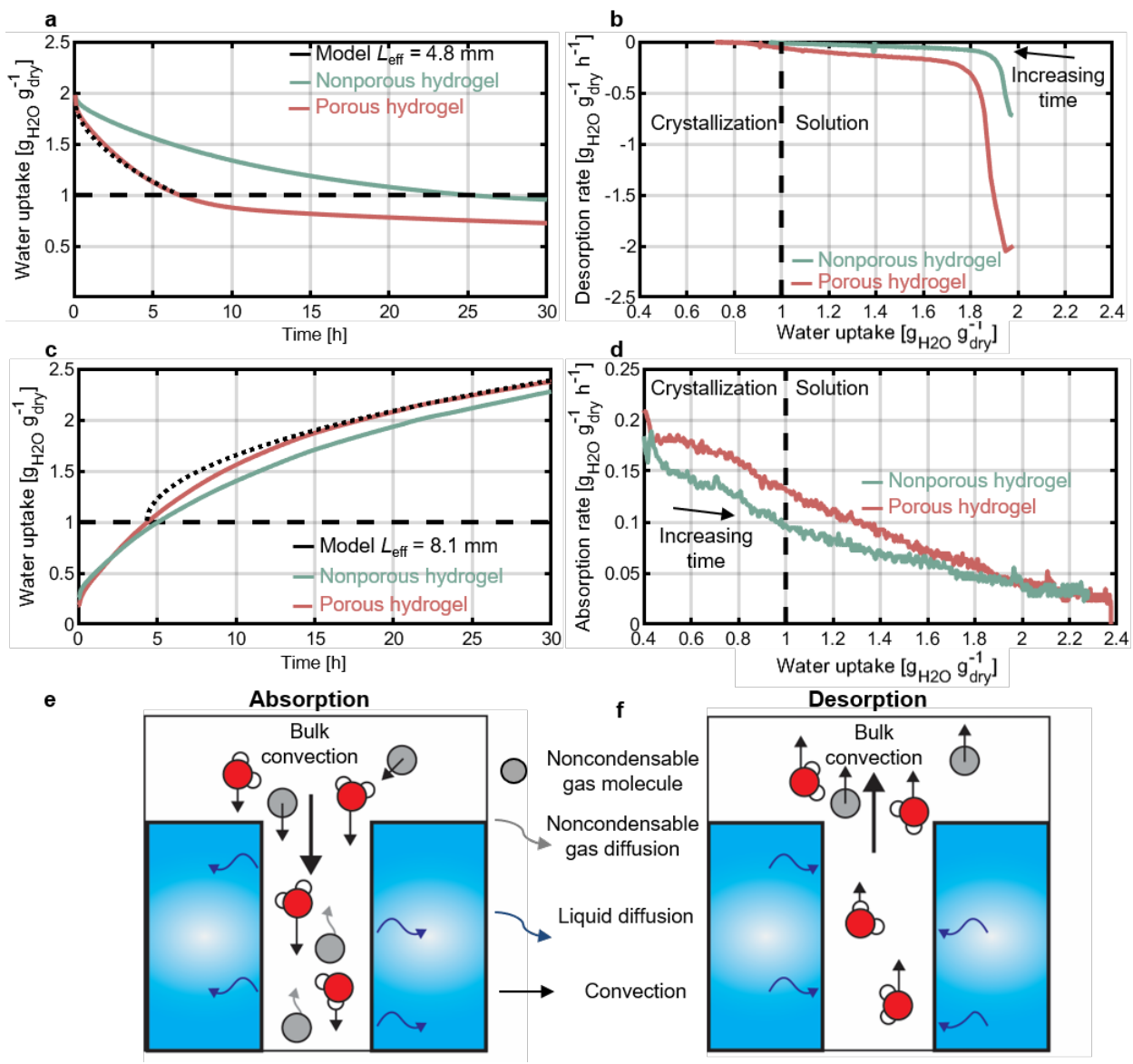

**Figure 4. Desorption and absorption of porous hydrogels. a**, Desorption of porous hydrogel is faster than that of the nonporous one, consistent with an effective diffusion length of $L_{eff}$ = 4.8 mm. **b**, When the salt is in a solution state, the desorption rate of the nonporous hydrogel is approximately 3 times higher than that of the nonporous sample. The desorption rates are similar once the salt in the hydrogels crystallized. In contrast, absorption of a porous hydrogel is only slightly faster than for a nonporous hydrogel as seen by, **c**, the uptake, which is consistent with $L_{eff}$ = 8.1 mm, and, **d**, the absorption rate. **e**, Absorption for a porous hydrogel is strongly affected by noncondensable gases. Both vapor and noncondensable molecules are convected to the hydrogel surface. The pores slow down diffusion of the noncondensables away from the surface, blocking vapor from condensing. **f**, During desorption, bulk convection is directed away from the pores. Therefore, there is no concentration of noncondensables.

**Figures 4c-d** compare the absorption of the nonporous and porous samples. When the uptake is lower than the crystallization condition, both samples have similar kinetics, as expected from the absence of significant water diffusion in the samples. Once the samples absorb enough water for the salt to deliquesce, the nonporous sample exhibits slightly faster absorption rates. We considered $D_{abs} = 1.81 \times 10^{-10}$ m² s⁻¹ (as obtained from the absorption of the nonporous sample) and fitted the experimental results to Equation (4) to obtain $L_{eff}$ = 8.1 mm. This value is considerably higher than that of desorption and it approaches the thickness of the sample, reflecting the smaller enhancement of kinetics during absorption. This difference in kinetics can be attributed to the prominent effect of noncondensable gases (NCGs), which are gas molecules different from water that have remained in the chamber. During absorption, there is bulk convection of vapor and NCGs towards the hydrogel surfaces (**Figure 4e**). While the vapor molecules condense, the NCGs concentrate at the hydrogel surface. In particular, for the long, narrow pores that we have designed, the concentration of NCGs will be significant as there is a strong resistance for these molecules to diffuse out of the pores. As a result, these NCGs will present an important transport barrier, blocking water vapor molecules from reaching the hydrogel surface. This result is unique to absorption. During desorption, the bulk convection has an opposite direction and it carries away both water molecules and NCGs, preventing NCG accumulation near the surface (**Figure 4f**). These observations are consistent with previous experiments of evaporation and condensation in pure-vapor environments[26–28] and explain their prominent effect during absorption of the porous samples. The nonporous samples do not exhibit differences in absorption and desorption kinetics

as it is easier for NCGs to diffuse away from the flat surface. Therefore, our experiments reveal the relevance of mitigating the non-condensable effect when operating in a pure-vapor environment, which is a relevant condition for thermal energy storage systems.[29,30] Despite the NCG effect, the strategy introduced here was successful in reducing the desorption and absorption timescales ($L_{eff}^2/D$) by 72% and 19%, respectively, with only a minimal loss in hydrogel volume of ~4%, demonstrating an optimized strategy towards performance enhancement in pure vapor environments. For systems operating in air-vapor mixture conditions, such as water harvesting and cooling systems, the effect of convection will lead to smaller reductions of the timescales.

In summary, we have probed the intrinsic water transport properties of hydrogel-salt composites. By combining experiments with macroscopic samples in pure-vapor conditions and modeling, we have measured for the first time the diffusion coefficient of water in a hydrogel-salt composite. Furthermore, by fabricating samples with rationally designed pores, we were able to significantly enhance the absorption and desorption kinetics. These experiments also provide key insights related to water diffusion in hydrogels and salt crystallization as well as on the effect of non-condensable gases in absorption. Altogether, these results advance our fundamental understanding of moisture-capturing hydrogels, which are critical towards material-level optimization across sorption applications.

ASSOCIATED CONTENT

**Supporting Information**.

The Supporting Information is available free of charge on the ACS Publications website at DOI: XXX (PDF)

AUTHOR INFORMATION

**Corresponding Author**

*Email: gustav.graeber@hu-berlin.de

**Author Contributions**

G.G. and C.D.D. contributed equally to this work. G.G. and C.D.D. conceived the initial concept. L.C.G. and G.G. synthesized samples and carried out sorption experiments. C.D.D. performed the modeling. G.G. and C.D.D wrote the manuscript with input from all coauthors.

**The authors declare no competing financial interest.**

ACKNOWLEDGMENT

The authors thank their advisor, Prof. Evelyn N. Wang, for her guidance on this work until the date of her confirmation (12/22/2022) by the U.S. Senate to serve as director of the Advanced Research Projects Agency-Energy (ARPA-E). The authors thank the Office of Energy Efficiency and Renewable Energy for funding under Grant No. DE-EE0009679. G.G. acknowledges funding by the Swiss National Science Foundation via a Postdoc Mobility grant (P400P2_194367). C.D.D. gratefully acknowledges the MIT Martin Family Society of Fellows for Sustainability. We thank Dr. Geoffrey Vaartstra, Dr. Bachir El Fil and Mr. Yang Zhong for fruitful discussions.